\documentclass[prd,twocolumn,preprintnumbers,amsmath,amssymb,nofootinbib,showpacs]{revtex4}
\usepackage{graphicx}
\usepackage{amssymb}
\newcommand{\beq}{\begin{equation}}
\newcommand{\eeq}{\end{equation}}
\usepackage{upgreek}
\usepackage{color}

\begin{document}

\title{Towards a gravity dual of the unitary Fermi gas} 
\date{\today} 
\author{Xavier Bekaert${}^1$, Elisa Meunier${}^1$ and Sergej Moroz${}^{2,3}$}

\affiliation{
\it ${}^1$ LMPT, Universit\'e Fran\c{c}ois Rabelais\\
UMR $7350$ CNRS \& FdR $2964$ Denis Poisson\\
 Parc de Grandmont 37200 Tours, France\\
\it ${}^2$ ITP, Universit{\"a}t Heidelberg\\
Philosophenweg 16, D-69120 Heidelberg, Germany\\
\it ${}^3$Department of Physics, University of Washington \\
Seattle, WA 98195-1560, USA\\
}

\begin{abstract} 
Inspired by the method of null dimensional reduction and by the holographic correspondence between Vasiliev's higher-spin gravity and the critical $O(N)$ model, a bulk dual of the unitary and the ideal non-relativistic Fermi gases is proposed.
\end{abstract}

\pacs{11.25.Tq, 03.75.Ss}

\maketitle

\section{Introduction} The quantum many-body problem of a non-relativistic two-component Fermi gas with short-range attractive interactions is a longstanding problem in condensed matter physics. At low temperature, the system is known to be superfluid and undergoes a smooth crossover from the Bardeen-Cooper-Schrieffer (BCS) to the Bose-Einstein-Condensate (BEC) regime as the two-body attraction is increased (see \cite{coldatomsreview} for reviews). Recent progress in experimental atomic physics allowed to study the BCS to BEC crossover with unprecedented accuracy.  The regime in between BCS and BEC, known as the ``unitary Fermi gas'', is of special theoretical interest. 
%In three spatial dimensions, 
The unitary Fermi gas is strongly coupled and no obvious small parameter is available precluding the reliable application of a perturbative expansion.

A characteristic of the unitary Fermi gas in vacuum is its invariance under scale transformations and, more generally, under the Schr\"odinger group of \cite{NiedererHagen}. This non-relativistic conformal symmetry of the unitary Fermi gas allowed \cite{Adscoldatoms} to apply the methods of gauge-gravity duality to this system. While these seminal papers triggered an intensive search for the holographic duals of various non-relativistic systems originating from condensed matter physics, a holographic description of the unitary fermions still remains tantalising. In this work, inspired by the conjectured anti de Sitter (AdS) dual of the critical $O(N)$ model \cite{Klebanov:2002ja}, we make a next step towards the gravity 
dual description of the unitary Fermi gas. 

\section{Unitary Fermi gas and its large-$N$ extension} Experimentally, a dilute two-component Fermi gas can be cooled with lasers to ultra-low temperatures close to absolute zero. Theoretically, this system can be very well described as a Fermi gas with two-body contact interactions governed by the microscopic BCS action
\begin{eqnarray} 
S[\,\psi\,;c_0]=\int dt \int d {\bf x}\, &\Big{[} \sum\limits_{\alpha\,=\,\uparrow,\downarrow}\psi_\alpha^*\left(i\partial_t +\frac{\Delta}{2m}+\mu \right)\psi_\alpha\, \nonumber\\ &\,\,-\,c_0\, \psi^*_\downarrow \psi^*_\uparrow\psi_\uparrow\psi_\downarrow \Big{]},\label{ac1}
\end{eqnarray}
where the two species of fermionic atoms of mass $m$ are represented by the Grassmann %-odd
fields $\psi_\uparrow$ and $\psi_{\downarrow}$, the chemical potential by $\mu$, and $c_0$ measures the strength of the interaction. This model has an internal $U(2)$ symmetry. Due to the contact nature of the interaction term, in three spatial dimensions ($d=3$) the quantum field theory defined by the action \eqref{ac1} must be renormalized by trading the bare interaction parameter $c_0$ for a low-energy observable: the s-wave scattering length.
Experimentally, the scattering length can be tuned via a Feshbach resonance by applying an external magnetic field. The unitary regime corresponds to an infinite scattering length.
Hence, in vacuum (\textit{i.e.} $\mu=0$) there is no intrinsic length scale in the unitary regime and the microscopic action \eqref{ac1} is invariant under the Schr\"odinger symmetry.
Most remarkably, the quantum version of this theory is believed to be an example of a strongly interacting non-relativistic conformal field theory (NRCFT) \cite{NRCFT}.

Aiming at a semi-classical holographic description, some large-$N$ extension of the unitary Fermi gas is necessary. A sensible construction that preserves the pairing structure of the interaction term was found in \cite{largeN}. The model with $N$ ``flavors'' is defined by the action
\beq \label{ac2}
\begin{split}
S[\,\psi\,;c_0,N]=\int dt \int d {\bf x}\, &\Big{[} \psi^\dagger\left(i\partial_t+\frac{\Delta}{2m}+\mu \right)\psi \\
&\,\,-\frac{c_0}{4N}\, \big|\,\psi^T \mathbb{J}\, \psi\,\big|^2 \Big{]},
\end{split}
\eeq
where $\psi$ denotes a multiplet of $2N$ massive fermions with components $\psi^A=\psi_\alpha^a$ with $\alpha=\,\uparrow,\downarrow$ and $a=1,\dots, N$. The symbol $\mathbb{J}$ represents the symplectic $2N\times2N$ matrix $\mathbb{J}_{AB}=\epsilon^{\alpha\beta}\otimes\delta_{ab}$. For $N=1$, one recovers the original model \eqref{ac1}. The extended model has $U(1)\times Sp\,(2N)$ as an internal symmetry group. Its subgroup $U(2)\times O(N)$, where $U(2)$ and $O(N)$ transform independently the ``spin'' and ``flavor'' indices (respectively $\alpha,\beta$ and $a,b$) will be central in our proposal.
Via a Hubbard-Stratonovich transformation in the Cooper channel, both theories \eqref{ac1}-\eqref{ac2} can be reformulated as effective field theories in terms of a complex scalar field (called dimer in the literature on the unitary Fermi gas) associated with the Cooper pair $\psi^T \mathbb{J}\, \psi\sim \psi_\uparrow\cdot\psi_\downarrow$.
In the large-$N$ limit the quantum partition function of \eqref{ac2} is equivalent to the mean field approximation for \eqref{ac1}, as both correspond to the saddle point of the dimer effective theory.

\section{Unitary Fermi gas \textit{vs} relativistic $U(N)$ model}
Relativistic scalar $O(N)$ models and their natural complex $U(N)$ extension are well understood due to their central role in the physics of critical phenomena. Despite their different space-time symmetries, the relativistic $U(N)$ model in $D=d+2$ dimensional space-time and the non-relativistic BCS model in $d$ spatial dimensions have several features in common. Some of the properties of the models are compared in the following table:
\begin{center}
\begin{tabular}{|c|c|c|}
 \hline
Models & $U(N)$ & BCS\\ 
\hline\hline
Space-time & Relativistic & Non-relativistic\\\hline
Fundamental fields & Bosons $\phi$ & Fermions $\psi_{\uparrow}$\,, $\psi_{\downarrow}$ \\\hline
Components & $N$ complex & $2N$ complex \\\hline
Internal symmetry & $U(N)$ & $U(1)\times Sp(2N)$ \\\hline
Quartic interaction & $(\phi^\dagger\cdot\phi)^2$ & $|\psi_\uparrow\cdot\psi_\downarrow|^2$\\\hline
Collective field & Particle density & Cooper pair \\ 
 & $\phi^\dagger\cdot\phi$ & $\psi_\uparrow\cdot\psi_\downarrow$ \\\hline
%Dimension & $D=d+2$ & $d=D-2$ \\\hline
Scale-free & $m=0$ & $\mu=0$ \\\hline
Critical fixed point & Wilson-Fisher & Unitary regime \\
\hline
%Kinematical sym. & $\mathfrak{o}(d+2,2)$ & $\mathfrak{sch}(d)$ \\\hline
%Higher-spin sym. & Vasiliev & Weyl\\\hline
\end{tabular}
\end{center}

$U(N)$ and BCS models have a similar renormalization group topology exhibiting a pair of fixed points. Besides the trivial fixed point, both theories can be tuned to criticality: the Wilson-Fisher fixed point for the massless $U(N)$ model and the unitary fixed point for the BCS model at $\mu=0$. In the large-$N$ limit, the models at the interacting fixed point are simply related to their non-interacting counterparts. In particular, by applying the general observation of \cite{Gubser:2002vv} to non-relativistic fermions, one can show that, in the large-$N$ limit, the free energies of the ideal and the unitary Fermi gases are related by a Legendre transformation with respect to the dimer field. Consequently, in this limit the theory at the two fixed points have the same infinite set of conserved currents and symmetries, most of which are  broken by $1/N$ corrections in the interacting theory. 
Analogous observation also holds for the relativistic $U(N)$ model. 

One also observes a simple relation between the scaling dimensions of the collective field at the two fixed points for both BCS and $U(N)$ models \cite{Adscoldatoms,Klebanov:2002ja}:
\begin{eqnarray} 
\Delta^{\mbox{free}}=\left\{ \begin{array}{c} d \qquad \text{BCS model}   \\ D-2 \qquad \text{$U(N)$ model} \\  \end{array} \right.
\nonumber\\
\Delta^{\mbox{int}}=2\quad\left\{ \begin{array}{c} \text{BCS model (in vacuum)}   \\ \text{$U(N)$ model ($N=\infty$ limit)} \\  \end{array} \right.
\label{scal}
\end{eqnarray}
In contrast to the relativistic $U(N)$ model, due to the simplicity of the non-relativistic vacuum, the relation $\Delta^{\mbox{int}}=2$ is exact in the theory of non-relativistic fermions, \textit{i.e.} it receives no $1/N$ corrections.

The highest of the two scaling dimensions, denoted $\Delta_+$, is always above the unitarity bound and corresponds to an infrared (IR) stable fixed point on the boundary side and to a standard (Dirichlet) boundary condition on the bulk side. The lowest dimension, $\Delta_-$, corresponds to an ultraviolet (UV) stable fixed point and to an exotic (Neumann) boundary condition.
Thus the holographic dual of the boundary Legendre transformation is a change of boundary conditions on the bulk scalar field.
When both dimensions are above the non-relativistic unitarity bound, $\Delta_+\geqslant \Delta_-\geqslant d/2$, both fixed points are admissible and thus correspond to different choices of boundary conditions for the \textit{same} bulk theory.

The unitary fixed point corresponding to $\Delta^{\mbox{int}}$ is physically admissible only for $0<d<2$ (IR stable) and $2<d<4$ (UV stable).
Indeed, for $d>4$ one obtains $\Delta^{\mbox{int}}=2<\frac{d}{2}$ which violates the unitarity bound for dimers. Moreover, in $d=2$ both fixed points merge together ($\Delta^{\mbox{free}}=2=\Delta^{\mbox{int}}$) and only the trivial fixed point exists.
The situation can be summarised as:
\begin{center}
\begin{tabular}{|c|c|c|c|}
 \hline
$d$ & $\Delta_-$ & $\Delta_+$ & Property\\ 
\hline\hline
%$d<0$ & - & $\Delta^{\mbox{int}}$ & only single admissible fixed pt \\\hline
%$d=0$ & $\Delta^{\mbox{free}}$ & $\Delta^{\mbox{int}}$ & saturation of unitarity bound \\\hline
$0<d<2$ & $\Delta^{\mbox{free}}$ & $\Delta^{\mbox{int}}$ & Asymptotic freedom
\\\hline
$d=2$ & 2 & 2 & Triviality \\\hline
$2<d<4$ & $\Delta^{\mbox{int}}$ & $\Delta^{\mbox{free}}$ & Asymptotic safety \\\hline
%$d=4$ & $\Delta^{\mbox{int}}$ & $\Delta^{\mbox{free}}$ & saturation of unitarity bound\\\hline
\end{tabular}
\end{center}

\section{Null dimensional reduction}
This is an old trick relating mathematically, relativistic and non-relativistic theories at tree level (see \textit{e.g.} \cite{Duval:2008jg}). It is based on the observation that the d'Alembertian of $D=d+2$ dimensional Minkowski space-time expressed in light-cone coordinates $x^\mu=(x^+,x^-,x^i)$ is proportional to the Schr\"odinger operator in $d$ spatial dimensions, modulo the identification of the null coordinate $x^+$ with the non-relativistic time and of the null momentum $-i\partial_-$ with the mass operator.
Indeed, the kinetic operator of a relativistic scalar theory, $\Box -M^2= -2\partial_+\partial_- + \Delta-M^2$, when acting on eigenmodes of the null momentum,
\beq\label{eigenm}
\Psi(x) = e^{-imx^-}\psi(x^+,x^i),
\eeq
is proportional to the kinetic Schr\"odinger operator of a non-relativistic theory , $i\partial_t+ \Delta/2m+\mu$, via the identification $x^+=t$ and $M^2=-\mu/2m$. %This follows from: 
%$$\big(\Box -M^2\big)\Psi(x)=2m\,e^{-imx^-}\Big(i\partial_+ + \frac{\Delta-M^2}{2m}\Big)\,\psi(x^+,x^i)\,.$$
Hence, any solution $\Psi(x)$ of the free Klein-Gordon equation of the form \eqref{eigenm} 
defines a solution $\psi(t,\textbf{x})$ of the free Schr\"odinger equation, and conversely. 

By definition, symmetries map solutions on solutions, thus the symmetries of the free Schr\"odinger equation can be seen as those symmetries of the free Klein-Gordon equation that commute with a fixed null momentum.
For instance, the Bargmann group (the central extension of the Galilei group by the mass)
and the Schr\"odinger group (the Bargmann group enlarged by expansions and scale transformations \cite{NiedererHagen})
are respectively the kinematical symmetry groups of the free Schr\"odinger equation with and without chemical potential \cite{Note}. They can be viewed as the centralisers of a given null momentum inside, respectively, the Poincar\'e and the conformal group of kinematical symmetries of the Klein-Gordon equation with and without mass.

The dimensional reduction explains the similarities between the large-$N$ critical $U(N)$ model and the unitary Fermi gas in vacuum. In fact, generally \emph{any} Lagrangian invariant under global $U(1)$ phase and Poincar\'e (conformal) transformations can be consistently reduced to a Lagrangian preserved by the $U(1)$ and Bargmann (Schr\"odinger) groups. 
This universal relationship between relativistic and non-relativistic field theories 
in the semi-classical (\textit{i.e.} mean field) approximation has maybe not yet received the attention that it deserves.

%For obtaining the large-$N$ extension of the unitary Fermi gas one can start with a free relativistic Grassman scalar theory in $D=d+2$ dimensional space-time and perform a null dimensional reduction. This generates a non-relativistic theory of free fermions in $d$ spatial dimensions. In the large-$N$ limit, the free energies of the ideal and the unitary Fermi gases are just Legendre conjugates of each other. Thence in this limit the results obtained from the null dimensional reduction of the free relativistic theory are of direct relevance for the challenging regime at unitarity. 

\section{Non-relativistic higher-spin symmetries}
A key feature of free CFTs is that their symmetries are enhanced to an infinite-dimensional higher-spin symmetry algebra. Following the holographic dictionary, the associated infinite
collection of higher-spin conserved currents should be dual to a tower of higher-spin gauge fields in the bulk. In particular, the bilinear singlet sector of a free scalar CFT should be dual to a Vasiliev theory \cite{Sezgin:2002rt}. 
Consequently, a crucial step towards a bulk dual of the ideal Fermi gas is the identification of symmetries and currents of the non-relativistic free fermions as well as their relationship to their relativistic parent. This lengthy analysis will be presented in details in \cite{Toappear} and here we only summarise our main results.

A target is the non-relativistic counterpart of the theorem of Eastwood \cite{Eastwood:2002su} identifying the maximal symmetry algebra of the d'Alembert equation in $D=d+2$ flat space-time. The latter infinite-dimensional algebra is denoted here as ``Vasiliev(d+2,2)'', since it contains the conformal algebra $\mathfrak{o}(d+2,2)$ and is used by Vasiliev as gauge algebra in his bosonic higher-spin theories on $AdS_{d+3}$ \cite{Vasiliev:2003ev}.
Mimicking the definitions of \cite{Eastwood:2002su},
a symmetry generator of the free Schr\"odinger equation (with $\mu=0$ from now on), 
\beq\label{freeSch}
\left(i\partial_t+ \frac{\Delta}{2m}\right)\psi(t,\textbf{x})=0\,\Leftrightarrow\, \psi(t,\textbf{x})=e^{it\frac{\Delta}{2m}}\psi(0,\textbf{x})\,,
\eeq
is a linear differential operator $\hat{A}$ such that $(i\partial_t+ \Delta/2m)\hat{A}=\hat{B}(i\partial_t+ \Delta/2m)$ for some linear differential operator $\hat{B}$, because then $\hat{A}$ maps solutions on solutions. 
Two generators are equivalent if they only differ by a trivial generator of the form
$\hat{A}=\hat{C}(i\partial_t+ \Delta/2m)$ for some linear operator $\hat{C}$, \textit{i.e.} $\hat A$ then maps solutions on zero. 
The maximal symmetry algebra of the free single-particle Schr\"odinger equation is the Lie algebra of all inequivalent symmetry generators and it is \cite{Toappear}:

$\bullet$ isomorphic to the Weyl algebra \cite{Vasiliev:2001zy}, denoted ``Weyl(d)'', of spatial differential operators (\textit{i.e.} quantum observables that are polynomials in positions and momenta) evolved in the time-reversed Heisenberg picture
$$
\hat{A}(-t)\,=\,e^{it\frac{\Delta}{2m}}\,\hat{A}(0)\,e^{-it\frac{\Delta}{2m}}
$$
that manifestly maps any solution \eqref{freeSch} to a solution,

$\bullet$ %can be 
generated algebraically by (taking powers of) the spatial translation and Galilean boost generators, $\hat{P}_i=-i \partial_i=\hat{P}_i(-t)$ and $\hat{K}_i=m x_i+i t \partial_i=m\hat{X}_i(-t)$ with canonical commutation relations $[\hat{K}_i,\hat{P}_j]=i\delta_{ij}m$,

$\bullet$ embedded in the Vasiliev algebra as the subalgebra commuting with a given null momentum and contains the Schr\"odinger algebra $\mathfrak{sch}(d)$, as summarised here:
\[%
\begin{array}
[c]{c|ccc}%
 & \mbox{Kinematical} & \subset& \mbox{Higher symmetries}\\
%&&&\\
\hline
\mbox{CFT} & \mathfrak{o}(d+2,2) & \subset & \mbox{Vasiliev (d+2,2)}\\
%&&&\\
\cup&\cup &  & \cup\\
%&&&\\
\mbox{NRCFT}&\mathfrak{sch}(d) & \subset & \mbox{Weyl (d)}
\end{array}
\]
where the vertical embedding corresponds to the null dimensional reduction, and the horizontal embedding  arises from the fact that the generators of kinematical symmetries are first-order differential operators while higher symmetries generators can be higher-derivatives. Notably \cite{Valenzuela:2009gu}, the Schr\"odinger algebra is contained in the Weyl algebra because its generators can be realized as polynomials of degree two in the spatial translation and Galilean boost generators.

For an $n$-component scalar field, these higher-spin space-time symmetry algebras of the d'Alembert and Schr\"odinger equations can be tensored with an internal $\mathfrak{u}(n)$ algebra of Hermitian $n\times n$ matrices. The corresponding higher-spin theories then possess $\mathfrak{u}(n)$-valued gauge fields \cite{Vasiliev:2003ev} dual to boundary 
bilinear currents in the adjoint representation of $U(n)$ \cite{Klebanov:2002ja}.

\section{Fermion bilinears and coupling to sources} The physical ($N=1$) BCS fermions are two-component Grassmann scalars in the fundamental representation of the internal symmetry group $U(2)$.
%The BCS theory lesson is that superfluidity and superconductivity have their origin in the condensation 
%of Cooper pairs. 
Together with the up and down particle densities, the Cooper pair fits into an adjoint multiplet of $U(2)$, \textit{i.e.} the $2\times2$ Hermitian matrix: $$
\left(
\begin{array}{cc} 
-\texttt{j}^{(0)}_{\,\uparrow} & \texttt{k}^{(0)}\\
\texttt{k}^{(0)*} & \texttt{j}^{(0)}_{\,\downarrow}
\end{array} 
\right):=\left(
\begin{array}{cc} 
-\psi^*_\uparrow\cdot\psi_\uparrow & \psi_\uparrow\cdot\psi_\downarrow\\
\psi^*_\downarrow\cdot\psi^*_\uparrow & \psi^*_\downarrow\cdot\psi_\downarrow
\end{array} 
\right)\,.
$$
In the large-$N$ extended theory, these considerations lead us to focus on the sector of flavor-singlet two-fermion composite fields in the adjoint representation of $U(2)$. They are spanned by the
$U(1)$-neutral conserved currents \cite{Gelfond:2008ur} (no sum over the index $\alpha=\uparrow,\downarrow$)
\beq\label{neutr}
\texttt{j}^{(r)}_{\alpha\,\,i_1 \cdots i_s}=\delta_{ab}\,\psi^{a*}_\alpha\,\underbrace{\overleftrightarrow{\partial_t}\cdots\overleftrightarrow{\partial_t}}_r\,\overleftrightarrow{\partial_{i_1}}\cdots\overleftrightarrow{\partial_{i_s}}\,\psi_\alpha^b
\eeq
and the $U(1)$-charged symmetric tensors
\beq\label{charg}
\texttt{k}^{(r)}_{i_1 \cdots i_s}=\delta_{ab}\,\psi^{a}_\uparrow\,\underbrace{\overleftrightarrow{\partial_t}\cdots\overleftrightarrow{\partial_t}}_r\,\overleftrightarrow{\partial_{i_1}}\cdots\overleftrightarrow{\partial_{i_s}}\,\psi_\downarrow^b\,.
\eeq
For $s=r=0$, these composite fields respectively reproduce the up and down particle densities $\texttt{j}_\alpha^{(0)}=\psi^*_\alpha\cdot\psi_\alpha$ and the Cooper pair $\texttt{k}^{(0)}=\psi_\uparrow\cdot\psi_\downarrow$.
%\textbf{Coupling to sources.} 
In the holographic correspondence, the composite operators \eqref{neutr}-\eqref{charg} should couple minimally to sources, respectively denoted by $h_{\alpha\,\,i_1\cdots i_s}^{(r)}$ and
$\upvarphi_{i_1\cdots i_s}^{(r)}$, representing the boundary data of $\mathfrak{u}(2)$-valued bulk gauge fields.
With the techniques of \cite{Bekaert:2010ky}, the difference of the free action $S[\,\psi\,;0,N]$ and of 
the minimal coupling term
$$
\sum\limits_{r,s\geqslant0}
\int dt\, d {\bf x}\,(j^{(r)i_1\cdots \,i_s}h_{i_1\cdots\, i_s}^{(r)}+k^{(r)i_1\cdots i_s*}\upvarphi_{i_1\cdots i_s}^{(r)}+\mbox{c.c.})\,,
$$
can be rewritten as the quadratic functional \cite{Toappear}
$$
\int dt\, d {\bf x}\,\, \Uppsi^\dagger\,\left(
\begin{array}{cc} 
i\partial_t+ \frac{\Delta}{2m}-\hat{H}_\uparrow & \hat{\upvarphi}\\
\hat{\upvarphi}^\dagger & i\partial_t- \frac{\Delta}{2m}+\hat{H}_\downarrow^\tau
\end{array} 
 \right)\,\Uppsi\, ,$$
where $\Uppsi^T\,=\,(\psi_\uparrow\,, \psi^{*}_\downarrow)$ defines the two-component Nambu-Gorkov fermion, $\hat{H}_\alpha^\tau(\hat{\mathbf{X}},\hat{\mathbf{P}}):=\hat{H}_\alpha(\hat{\mathbf{X}},-\hat{\mathbf{P}})$
and the differential operators $\hat{H}_\alpha$ and $\hat{\upvarphi}$ are related to the respective sources $h_\alpha$ and
$\upvarphi$. 
This compact rewriting is formally identical to the Nambu-Gorkov formulation of the BCS theory except that the chemical potential and the energy gap are replaced by space-time differential operators.
The effective action can be obtained now via a Gaussian integration over the fermionic field and is a Trace-log functional of the above $2\times 2$ matrix. These results can be reproduced through the null dimensional reduction of a free relativistic scalar theory \cite{Toappear}.

\section{Bulk dual}
\textit{What might be the gravity dual of the unitary Fermi gas?} Keeping the above discussion in mind, we approach this question by following these steps: (i) unitary fermions at $N=\infty$ are Legendre conjugate to free fermions, (ii) a key tool for understanding higher-spin symmetries of free non-relativistic fermions is the null dimensional reduction of free relativistic Grassmann scalars, (iii) free relativistic scalar theories are expected to be dual to Vasiliev higher-spin theories.

Therefore it is tempting to perform the null reduction on both sides of the relativistic holographic duality, as in \cite{ads}.
Schematically, our philosophy looks as follows:
\begin{center}
\begin{tabular}[t]{ccc}
Higher-spins on AdS$_{d+3}$ & $\longleftrightarrow$ & CFT$_{d+2}$\\
$\downarrow$ &  & $\downarrow$ \\
Non-relativistic Higher-spins$_{d+1}$ & $\overset{?}{\longleftrightarrow}$ & NRCFT$_{d}$
\end{tabular}
\end{center}
with horizontal arrows denoting holographic correspondence and vertical arrows relating relativistic to non-relativistic theories via the null reduction.
The higher-dimensional relativistic parents are mere auxiliary tools in our construction, used at tree level only since they may be sick as quantum field theories \textit{per se}. For instance, the CFT violates the spin-statistics theorem, but this is not a problem since this theorem does not apply to non-relativistic theories.

We thus propose that a candidate for the holographic description of fermions at unitarity is the null reduction of a Vasiliev higher-spin gravity \cite{NRG}. More precisely, the $O(N)$-invariant sector of the large-$N$ unitary fermions in $d$ spatial dimensions might be dual to the null reduction of the Vasiliev bosonic theory \cite{Vasiliev:2003ev} on $AdS_{d+3}$ with $U(2)$ internal symmetry. Scalar fields on $AdS_{d+3}$ admit two distinct choices of boundary conditions for mass-square in the interval $-(\frac{\,d+2}{\,2})^2< m^2< 1-(\frac{\,d+2}{\,2})^2$. Since the complex bulk scalar fields in the higher-spin multiplet have $m^2=-2d$, this possibility arises in the intervals $0<d<2$ and $2<d<4$.
In particular, the gravity dual of the ``physical'' three-dimensional ($d=3$) two-component ($N=1$) UV-stable ($\Delta_-=2$) unitary Fermi gas should be \emph{the null dimensional reduction of Vasiliev theory describing interacting $\mathfrak{u}(2)$-valued higher-spin gauge fields on $AdS_6$ with exotic boundary condition for the bulk scalar field dual to the Cooper pair}.
The intimate connection between the unitary and the ideal Fermi gases together with the universality of the null dimensional reduction method suggest that the holographic dual of the unitary Fermi gas is within our immediate reach.

\textbf{Acknowledgments}
We thank C. Duval, S. Golkar, M. Hassaine, P. Horvathy, D.T. Son, J. Unterberger and M. Valenzuela for discussions on related issues. Work of S.M. was supported by U.S. DOE Grant No. DE-FG02-97ER41014.


\begin{thebibliography}{11}
% \providecommand{\natexlab}[1]{#1}
% \providecommand{\url}[1]{\texttt{#1}}
% \expandafter\ifx\csname urlstyle\endcsname\relax
%   \providecommand{\doi}[1]{doi: #1}\else
%   \providecommand{\doi}{doi: \begingroup \urlstyle{rm}\Url}\fi
\bibitem{coldatomsreview}
I.~Bloch, J.~Dalibard and W.~Zwerger,
  Rev.\ Mod.\  Phys {\bf 80} (2008) 885;
  %[arXiv:0704.3011 [cond-mat]];
S.~Giorgini, L.~P.~Pitaevskii and P~Stringari,
 Rev.\ Mod.\ Phys.\  {\bf 80} (2008) 1215.
 %[arXiv:0706.3360 [cond-mat]].

\bibitem{NiedererHagen}
  U.~Niederer, 
  %``The maximal kinematical invariance group of the free Schr\"odinger 
  % equation,'' 
  Helv.\ Phys.\ Acta {\bf 45} (1972) 802;
  C.~R.~Hagen,
  %``Scale and conformal transformations in Galilean-covariant field theory,''
  Phys.\ Rev.\  D {\bf 5} (1972) 377.
  %%CITATION = PHRVA,D5,377;%%

\bibitem{Adscoldatoms}
  D.~T.~Son,
  %``Toward an AdS/cold atoms correspondence: A Geometric realization of the
  %Schrodinger symmetry,''
  Phys.\ Rev.\  D {\bf 78} (2008) 046003;
  %[arXiv:0804.3972 [hep-th]].
  %%CITATION = PHRVA,D78,046003;%%
%\bibitem{Balasubramanian:2008dm}
  K.~Balasubramanian and J.~McGreevy,
  %``Gravity duals for non-relativistic CFTs,''
  Phys.\ Rev.\ Lett.\  {\bf 101} (2008) 061601.
  %[arXiv:0804.4053 [hep-th]].
  %%CITATION = PRLTA,101,061601;%%

\bibitem{Klebanov:2002ja}
  I.~R.~Klebanov and A.~M.~Polyakov,
  %``AdS dual of the critical O(N) vector model,''
  Phys.\ Lett.\  B {\bf 550} (2002) 213.
  %[arXiv:hep-th/0210114].
  %%CITATION = PHLTA,B550,213;%%

\bibitem{NRCFT}
  T.~Mehen, I.~W.~Stewart and M.~B.~Wise,
  %``Conformal invariance for non-relativistic field theory,''
  Phys.\ Lett.\  B {\bf 474} (2000) 145;
  %[arXiv:hep-th/9910025].
  F.~Werner and Y.~Castin,
  Phys.\ Rev.\ A {\bf 74} (2006) 053604;
%\bibitem{Nishida}
  Y.~Nishida and D.~T.~Son,
  %``Nonrelativistic conformal field theories,''
  Phys.\ Rev.\  D {\bf 76} (2007) 086004.
  %[arXiv:0706.3746 [hep-th]].
  %%CITATION = PHRVA,D76,086004;%%

\bibitem{largeN}
  P.~Nikolic and S.~Sachdev,
  %``Renormalization-group fixed points, universal phase diagram, and 1/N
  %expansion for quantum liquids with interactions near the unitarity limit,''
  Phys.\ Rev.\  A {\bf 75} (2007) 033608;
  %[arXiv:cond-mat/0609106];\\
  %%CITATION = PHRVA,A75,033608;%%
  %\bibitem{Veillette:2007zz}
  M.~Y.~Veillette, D.~E.~Sheehy and L.~Radzihovsky,
  %``Large-N expansion for unitary superfluid Fermi gases,''
  Phys.\ Rev.\  A {\bf 75} (2007) 043614.
  %[arXiv:cond-mat/0610798].
  %%CITATION = PHRVA,A75,043614;%%

\bibitem{Gubser:2002vv}
  S.~S.~Gubser and I.~R.~Klebanov,
  %``A Universal result on central charges in the presence of double trace
  %deformations,''
  Nucl.\ Phys.\  B {\bf 656} (2003) 23.
%  [arXiv:hep-th/0212138].
  %%CITATION = NUPHA,B656,23;%%

\bibitem{Duval:2008jg}
  C.~Duval, M.~Hassaine, P.~A.~Horvathy,
  %``The Geometry of Schrodinger symmetry in gravity background/non-relativistic CFT,''
  Annals Phys.\  {\bf 324 } (2009)  1158-1167.
%  [arXiv:0809.3128 [hep-th]].

\bibitem{Note}
Non-relativistic kinematical symmetries are symmetry transformations, $\psi(t,{\bf x})\to 
\gamma\big(g(t,{\bf x})\big) \,\psi\big(g^{-1}(t,{\bf x})\big)$, thus its generators are first-order differential operators.
    
\bibitem{Sezgin:2002rt}
  E.~Sezgin and P.~Sundell,
  %``Massless higher spins and holography,''
  Nucl.\ Phys.\  B {\bf 644} (2002) 303.
%  [Erratum-ibid.\  B {\bf 660} (2003) 403]
%  [arXiv:hep-th/0205131].
  %%CITATION = NUPHA,B644,303;%%  

\bibitem{Valenzuela:2009gu}
  M.~Valenzuela,
  %``Higher-spin symmetries of the free Schr\"odinger equation,''
  [arXiv:0912.0789 [hep-th]].
  %%CITATION = ARXIV:0912.0789;%%
  
\bibitem{Toappear}
  X.~Bekaert, E.~Meunier and S.~Moroz,
  %``Symmetries and currents of the ideal and unitary Fermi gases,''
  arXiv:1111.3656 [hep-th].
  %%CITATION = ARXIV:1111.3656;%%
      
\bibitem{Eastwood:2002su}
  M.~G.~Eastwood,
  %``Higher symmetries of the Laplacian,''
  Annals Math.\  {\bf 161} (2005) 1645.
%  [arXiv:hep-th/0206233].
  %%CITATION = ANMAA,161,1645;%%

\bibitem{Vasiliev:2003ev}
  M.~A.~Vasiliev,
  %``Nonlinear equations for symmetric massless higher spin fields in
  %(A)dS(d),''
  Phys.\ Lett.\  B {\bf 567} (2003) 139.
  %[arXiv:hep-th/0304049].
  %%CITATION = PHLTA,B567,139;%%
  
\bibitem{Vasiliev:2001zy}
  This result is a corollary of the following general results on $\mathfrak{sp}(2d,\mathbb R)$-covariant unfolded equations:
  M.~A.~Vasiliev,
  %``Conformal higher spin symmetries of 4-d massless supermultiplets and osp(L,2M) invariant equations in generalized (super)space,''
  Phys.\ Rev.\ D {\bf 66} (2002) 066006.
%  [hep-th/0106149].
  %%CITATION = HEP-TH/0106149;%%

\bibitem{Gelfond:2008ur}
The general structure of such non-relativistic higher-spin conserved currents was found in:
  O.~A.~Gelfond and M.~A.~Vasiliev,
  %``Higher Spin Fields in Siegel Space, Currents and Theta Functions,''
  JHEP {\bf 0903} (2009) 125.
%  [arXiv:0801.2191 [hep-th]].
  %%CITATION = ARXIV:0801.2191;%%

\bibitem{Bekaert:2010ky}
  X.~Bekaert, E.~Joung and J.~Mourad,
  %``Effective action in a higher-spin background,''
  JHEP {\bf 1102} (2011) 048.
%  [arXiv:1012.2103 [hep-th]].
  %%CITATION = JHEPA,1102,048;%%
  
\bibitem{ads}
  W.~D.~Goldberger,
  %``AdS/CFT duality for non-relativistic field theory,''
  JHEP {\bf 0903} (2009) 069;
  %[arXiv:0806.2867 [hep-th]];
  %%CITATION = JHEPA,0903,069;%%
  J.~L.~F.~Barbon and C.~A.~Fuertes,
  %``On the spectrum of nonrelativistic AdS/CFT,''
  JHEP {\bf 0809} (2008) 030;
  %[arXiv:0806.3244 [hep-th]];
  %%CITATION = JHEPA,0809,030;%%
  F.~L.~Lin and S.~Y.~Wu,
  %``Non-relativistic Holography and Singular Black Hole,''
  Phys.\ Lett.\  B {\bf 679} (2009) 65.
  %[arXiv:0810.0227 [hep-th]].
  %%CITATION = PHLTA,B679,65;%%
\bibitem{NRG}
Although the null reduction is tricky in gravity, a promising route is to follow:
C.~Duval, G.~Burdet, H.~P.~Kunzle and M.~Perrin, Phys.\ Rev.\  D {\bf 31} (1985) 1841; B.~Julia and H.~Nicolai, Nucl.\ Phys.\  B {\bf 439}, 291 (1995).

    
\end{thebibliography}
\end{document}